%% file: ppqesnew.tex
\documentclass[preprint,aps,pra]{revtex4}
\newcommand{\eqref}[1]{(\ref{#1})}
\def\sn{{\rm sn\,}}
\def\cn{{\rm cn\,}}
\def\dn{{\rm dn\,}}

\begin{document}

\title{\bf{Periodic Quasi-Exactly Solvable Models }}
\author{
  S. Sree Ranjani$^{a}$\footnote{akksprs@uohyd.ernet.in},
  A. K. Kapoor$^{a}$\footnote{akksp@uohyd.ernet.in} and
  P. K. Panigrahi$^{a,b}$\footnote{prasanta@prl.ernet.in}}
 \address{$^a$ School of Physics, University of Hyderabad, Hyderabad 500 046, India \\
$^b$ Physical Research Laboratory Navrangpura, Ahmedabad, 380009,
India}
\begin{abstract}
Various quasi-exact solvability conditions, involving the
parameters of the periodic associated Lam{\'e} potential, are
shown to emerge naturally in the quantum Hamilton-Jacobi approach.
It is found that, the intrinsic nonlinearity of the Riccati type
quantum Hamilton-Jacobi equation is primarily responsible for the
surprisingly large number of allowed solvability conditions in the
associated Lam{\'e} case. We also study the singularity structure
of the quantum momentum function, which yields the band edge
eigenvalues and eigenfunctions.
\end{abstract}

\maketitle

\section{Introduction}

Quasi-exactly solvable (QES) Hamiltonians, interconnecting a
diverse array of physical problems, have been the subject of
extensive study in recent times
[\ref{Singh},\ref{Turb},\ref{Shifm},\ref{Ushve},\ref{geo},\ref{Atre}].
These systems, containing a finite number of exactly obtainable
eigenstates, have been linked with classical electrostatic
problems, as also to the finite dimensional irreducible
representations of certain algebras. Some of these studies employ
group theoretical methods, others are based on the symmetry of the
relevant differential equations
[\ref{Turb},\ref{Shifm},\ref{Ushve}]. The key to the existence of
the finite number of identifiable states is the quasi-exact
solvability condition, relating certain potential parameters of
these dynamical systems. An interesting feature, distinguishing
these QES systems from known exactly solvable cases, is the
presence of complex zeros of the wave
functions. Hence, quantum Hamilton-Jacobi (QHJ) formalism, being
naturally formulated in the complex domain, is ideally suited for
studying the QES problems [\ref{geo}]. Although polynomial
potentials have been studied rather exhaustively, QES periodic
potentials have not received significant attention in the
literature.

     The associated Lam{\'e} potential (ALP)
\begin{equation}
V(x) = a(a+1)m \, \sn^2(x,m) + b(b+1) m \,
\frac{\cn^2(x,m)}{\dn^2(x,m)}, \label{e2}
\end{equation}
 is an interesting example of a periodic potential, which is
 exactly solvable, when $a=b$ and shows QES property, when $a\neq
 b$. Here, $\sn(x,m),\cn(x,m)$ and $\dn(x,m)$ are the doubly
 periodic elliptic functions with modulus parameter $m$ [\ref{han}].
The ALP has a periodic lattice of period $K(m)$ with the basis
composed of two different atoms which are alternately placed. It
possesses a surprisingly large variety of QES solvability
conditions depending on the nature of the potential parameters $a$
and $b$.

 For $a$, $b$ being unequal integers, with $a>b>0$,
there are $a$ bound bands followed by a continuum band. If $a-b$
is odd (even) integer, it has $b$ doubly degenerate band edges of
period $2K(4K)$, which can not be obtained analytically. For $a$,
$b$ having half-integral values (with $a>b$), there are infinite
number of bands with band edge wave functions having period $2K$
($4K$), if $a-b$ is odd (even). Of these infinite number of bands,
$ a-b $ bands have band edges, which are non-degenerate with
period $2K(4K)$ and $b+\frac{1}{2}$ doubly degenerate states of
period $2K(4K)$ which can be obtained analytically. For $a$ being
an integer and $b$ being a half integer or vice versa, one can
obtain  some exact analytical results for mid-band states
[\ref{khare}, \ref{suk}, \ref{mag}].

It is quite natural to enquire about the origin of this rich QES
structure in the associated Lam{\'e} potential
[\ref{asis},\ref{tka}]. This paper is devoted to the study of the
same through the quantum Hamilton-Jacobi approach. The quasi-exact
solvability conditions, involving the parameters of the periodic
associated Lam{\'e} potential, are shown to emerge naturally. The
intrinsic nonlinearity of the Riccati type QHJ equation is
responsible for these allowed solvability conditions in the
associated Lam{\'e} case. We also study the singularity structure
of the quantum momentum function, which yields the band edge
eigenvalues and eigenfunctions.

   In our earlier studies, we had looked at non-periodic ES, QES and
ES periodic potentials through the QHJ formalism which was
initiated by Leacock and Padgett [\ref{lea}, \ref{pad}]. We were
successful in obtaining the quasi exact solvability condition
[\ref{geo}] for QES models and in obtaining the eigenvalues and
eigenfunctions for both the ES [\ref{sree}] and QES models
[\ref{geo}]. Within the same formalism, we have also proposed a
new method to solve ES periodic potentials [\ref{akk}] and obtain
the band edge eigenvalues and eigenfunctions. The present study
makes use of the QHJ formalism as devolved in our earlier papers
[\ref{sree}, \ref{geo}] and [\ref{akk}]. We refer the interested
readers to them for all the details.

   The QHJ formalism revolves round the logarithmic derivative of
the wave function $p$,
\begin{equation}
p = -i\hbar \frac{d}{dx}\ln \psi   \label{e3}
\end{equation}
known as the quantum momentum function (QMF), which satisfies a
Riccati type equation
\begin{equation}
p^2 -i\hbar \frac{dp}{dx} = 2M(E-V(x)).     \label{e4}
\end{equation}
We have found that, the knowledge of the singularity structure of
the QMF and the residue at these singular points are the only
information  needed to obtain the QES condition and the
solutions. In all our earlier studies, it was assumed that, {\it
the point at infinity is an isolated singularity}, which turned
out to be true in all the cases.
    With this same assumptions on $p$, we proceed to obtain the QES
condition and the forms of the band edge wave functions for the
general ALP in the next section. In this paper we concentrate on
the cases where, $a$ and $b$ are either integers or half integers.
In section III, we analyze the situation where both $a$, $b$ are
integers taking, the values $2$ and $1$, respectively. The case
when $a$ and $b$ are both half integers, with values $7/2$ and
$1/2$ respectively, is analyzed in section IV.

\section{QES condition and the forms of the wave functions}
The QHJ equation for the ALP, putting $\hbar = 2M = 1$, is given
by
\begin{equation}
p^2 - ip^{\prime} = \left( E - a(a + 1)m \, \sn^2(x) - b(b + 1)m \,
  \frac{\cn^2(x)}{\dn^2(x)}\right).  \label{e5}
\end{equation}
Note that with the transformation $b \rightarrow - b - 1$ or $a
\rightarrow - a - 1$, the potential does not change. Hence, for
our analysis, without loss of  generality, we take $a,b$
positive, with $a > b$. Defining $p \equiv -iq$ and substituting
it in \eqref{e5}, one obtains
\begin{equation}
q^2 + \frac{dq}{dx} = a(a + 1)m \, \sn^2(x) + b(b + 1)m
\,\frac{\cn^2(x)}{\dn^2(x)} - E.   \label{e6}
\end{equation}
Changing the variable to $ t \equiv \sn (x) $, and writing
\begin{equation}
q = \sqrt{(1 - t^2)(1 - m \, t^2)}\phi, \quad {\mathrm{with}}
\quad \phi = \chi + \frac{1}{2}\left(\frac{m
  \,t}{1 - m \,t^2}+\frac{t}{1 - t^2}\right),     \label{e7}
\end{equation}
Equation \eqref{e6}, gets transformed into
\begin{equation}
\chi^2 + \frac{d\chi}{dt} + \frac{m^2t^2 + 2m(1-2b(b + 1))}{4(1 - mt^2)^2} +
\frac{2 + t^2}{4(1 - t^2)^2} + \frac{2E - mt^2(1 - 2a(a + 1))}{2(1 - t^2)(1 -
  mt^2)} = 0.      \label{e8}
\end{equation}
For all our later calculations, we shall treat $\chi$ as the QMF and
\eqref{e8} as the QHJ equation.

{\bf Singularity Structure : } From \eqref{e8}, we see that $\chi$
has fixed poles at $ t = \pm 1$ and $t = \pm \frac{1}{\sqrt{m}}$.
In addition to the fixed poles, the QMF has finite number of
moving poles and no other singular points in the complex plane.
Hence, one can write $\chi$ as a sum of the singular and
analytical parts as follows
\begin{equation}
\chi = \frac{b_1}{t - 1} + \frac{b_1^\prime}{t + 1} + \frac{d_1}{t -
  \frac{1}{\sqrt m}} + \frac{d_1}{t + \frac{1}{\sqrt m}} +
  \frac{P_n^\prime}{P_n} + Q(t) \, ,    \label{e9}
\end{equation}
where $b_1, b_1^\prime, d_1$ and $ d_1^\prime$ are the residues at $t = \pm 1$
and $\pm \frac{1}{\sqrt m}$ respectively, which need to be calculated. $P_n$
is an $n^{th}$ degree polynomial with
 $ \frac{P_n^\prime}{P_n} = \sum_{k=1}^n{\frac{1}{t-t_k}}$ 
being the summation  of terms coming from the $n$ moving poles
with residue one. The function $Q(t)$ is analytic and bounded at
infinity. Hence, from Liouville's theorem it is a constant, say
$C$. The residues at the fixed poles can be calculated by taking
the Laurent expansion around each individual pole and substituting
them in \eqref{e8}. Comparing the coefficients of different powers
of $t$, one gets two values of residues at each pole owing to the
quadratic nature of the QHJ equation. Thus the two values of the
residues, $b_1$ ,$ b^{\prime}_1 $ at $t = \pm 1$ are
\begin{equation}
b_1 = \frac{3}{4},\frac{1}{4}  \,\,\,\,\,\,\,\text{and}\,\,\,\, \,\,\,  b^{\prime}_1 = \frac{3}{4},\frac{1}{4}.  \label{e11}
\end{equation}
At $t = \pm \frac{1}{\sqrt m}$, one has
\begin{equation}
d_1 = \frac{3}{4} + \frac{b}{2},\frac{1}{4} - \frac{b}{2} \quad
\text{and} \quad d^{\prime}_1 = \frac{3}{4} + \frac{b}{2},
\frac{1}{4} - \frac{b}{2}.  \label{e12}
\end{equation}
    Since, there is no way of ruling out one of the two values of these
residues, we need to consider both the values. We demand $b_1 =
b^{\prime}_1$ and $ d_1 = d_1^\prime $, a condition coming from
the parity constraint $ \chi (t) = - \chi (t)$.


{\bf Behaviour at infinity : } Equation \eqref{e9} gives the
behaviour of $\chi$ in the entire complex plane. Hence, for large
$t$,
\begin{equation}
\chi \sim \frac{2b_1 + 2d_1 + n}{t} \, , \label{e13}
\end{equation}
 where the restriction $b_1 = b_1^\prime$ and $d_1 = d_1^\prime$ has been
 applied. This should match with the leading behaviour of $\chi$
 obtainable from  \eqref{e8}. Note that, the assumption on the
 singularity structure on  $\chi$ is equivalent to the point at
 infinity being an isolated
 singularity.  Since $\chi$ has at most an isolated singular point at
 infinity, one can expand
 $\chi$ in Laurent series around the point at infinity as,
\begin{equation}
\chi(t) = \lambda_0 +\frac{\lambda_1}{t} +\frac{\lambda_2}{t^2}
+\cdots \, \, . \label{e14}
\end{equation}
Substituting \eqref{e14} in \eqref{e8} and comparing various
powers of $t$, one gets $\lambda_0 = 0 $, thus making $Q(t)$ in
\eqref{e9} equal to zero. Further one obtains
\begin{equation}
\lambda_1 = a+1    ,\,\,\,  -a.    \label{e15}
\end{equation}
Since both the equations \eqref{e13} and \eqref{e15} give the
leading behavior of $\chi$ at infinity, both should be equal. Thus
\begin{equation}
2b_1 + 2d_1 + n =  \lambda_1.        \label{e16}
\end{equation}
Taking various combinations of $b_1$ and $d_1$ from \eqref{e11}
and \eqref{e12}, substituting them in \eqref{e16} one obtains the
QES condition for each combination as given in table 1 for
$\lambda = a+1$. Thus one sees that all the allowed combinations
of residues give one of the forms of QES condition [\ref{khare}],
where $n = 0, 1, 2 \cdots$ . Note that, the other value of
$\lambda$, {\it i.e.,} $-a$, when substituted instead of $a + 1$
in \eqref{e16}, gives the QES condition for negative values of $a,
b$ {\it i.e.,} for $a \rightarrow - a - 1,\, b \rightarrow -b - 1$
in $b_1,\,d_1$.

{\bf Forms of wave function :} From \eqref{e3}, one can write $\psi$
in terms of $p$ as
\begin{equation}
\psi(x) = \exp \left( \int{ip dx} \right).  \label{e17}
\end{equation}
Changing the variable to $t$ and writing $p$ in terms of $\chi$, one gets \begin{equation}
\psi(x) = \exp \left (\int{\left(\chi + \frac{1}{2}\left(\frac{mt}{1-mt^2} +
  \frac{t}{1-t^2}\right) \right)dt} \right).  \label{e18}
\end{equation}
Substituting  $\chi$ from \eqref{e9} in the above equation gives
the wave function in terms of the residue $b_1,\, d_1$ and the
polynomial $P_n$ :
\begin{equation}
\psi(t) =  \exp \left( \int{(\frac{(1-4b_1)t}{2(1-t^2)} + \frac{(1-4d_1)mt}{2(1-mt^2)} +
  \frac{P_n^\prime}{P_n})dt}. \right)  \label{e19}
\end{equation}
In terms of the original variable $x$ the wave function takes the
form,
\begin{equation}
\psi(x) = (\cn x)^\alpha (\dn x)^\beta P_n (\sn x)  \label{e20}
\end{equation}
where $\alpha = \frac{4b_1 - 1}{2}, \,\,\,  \beta = \frac{4d_1 -
1}{2} $. Hence, for each set of $b_1,\,d_1$ one gets a wave
function given by \eqref{e20}. The degree $n$ of this polynomial,
which is obtained from \eqref{e16} as,
\begin{equation}
n = a + 1 - 2b_1 - 2d_1    \label{e22}
\end{equation}
is in terms of either $a + b$ or $a - b$,  as evident from table
1. The forms of the wave function can be found and are given in
table 1, for the two different cases, when $a+b$ and $ a-b$ are
odd and
even separately.\\
{\bf Case 1}, {\it  Both $a+b$, $a-b$ are even}: We introduce $N =
\frac{a+b}{2}$ and $ M =\frac{a-b}{2}$, where $M$ and $N$ are
integers, and obtain the forms of the wave functions in table 2,
in terms of $M$ and $N$, for the four sets of combinations of
$b_1$ and $d_1$ in table 1.\\

 {\bf Case 2} {\it Both $a+b$, $a-b$ odd}: Introducing
$N^{\prime}=\frac{a+b}{2}$ and $M^{\prime} = \frac{a-b}{2}$, where
 $M^{\prime}$ and $N^{\prime}$ are integers, we obtain the wave
 functions, in terms of $M^{\prime}$ and $N^{\prime}$,
for the four sets of combinations of $b_1$ and $d_1$ in table 1.\\

   From the forms of the wave functions in tables 2 and 3, one observes
that the number of linearly independent solutions is different for
the two cases. The unknown polynomial in the wave function can be
obtained by substituting $\chi$ from \eqref{e9} in the QHJ
equation \eqref{e8}, which gives
\begin{equation}
P_{n}^{\prime\prime}(t) + 4P_{n}(t)\left(\frac{b_1 t}{t^2 -1} +
\frac{md_1 t}{mt^2 -1}\right)+G(t)P_{n}(t) =0    \label{e23}
\end{equation}
where
\begin{eqnarray}
G(t) =\frac{t^2 (4b_{1}^{2}-2b_1 +\frac{1}{4})-2b_1
+\frac{1}{2}}{(t^2 -1)^2} + \frac{m^2 t^2 (4d_1 ^2 -2d_1
+\frac{1}{4})- 2md_1 + m(\frac{1-2b(b+1)}{2})}{(mt^2 -1)^2} +
\nonumber \\
\frac{2E +(16b_1d_1 -1 -2a(a+1))mt^2}{2(1-t^2)(1-mt^2) }.\nonumber
\end{eqnarray}
 The above differential equation is equivalent to a system of $n$ linear equations for the
coefficients of the different powers of t in $P_n(t)$. The energy
eigenvalues are obtained by setting the corresponding determinant
equal to zero. In the next section we obtain the band edge wave
functions for the associated Lam{\'e} potential, when $a=2$ and $b
= 1$.

\section{ ALP with \protect\( $a, b$ \protect \) integers :} For this case, we consider the
associated Lam{\'e} potential with $a = 2,\,b = 1$. For the
purpose of comparison with literature, we work with the
supersymmetric potential
\begin{equation}
V_-(x) = 6m \,{\sn(x)}^2 + 2m \frac{{\cn x}^2}{{\dn x}^2}
  - 4m   \label{e24}
\end{equation}
This potential is same as \eqref{e2} with $a=2$ and $b=1$,
except that a constant has been added to make the
lowest energy equal to  zero. The QHJ equation in terms
of $\chi$ is
\begin{equation}
\chi^2 + \chi^{\prime} + \frac{m^2 t^2 - 6m}{4(1 - mt^2)^2} +
\frac{2 + t^2}{4(1 - t^2)^2} + \frac{2E +8m -13mt^2}{2(1 - t^2)(1-
mt^2)} = 0.      \label{e25}
\end{equation}
Apart from $n$ moving poles, $\chi$ has poles at $t=\pm 1$ and
$t=\pm 1/ \sqrt{m}$. As in the previous section one can write
$\chi$, with the parity constraint as
\begin{equation}
\chi = \frac{2b_1t}{t^2 -1} +\frac{2md_1t}{t^2 - \frac{1}{m}}+
\frac{P_{n}^{\prime}(t)}{P_n (t)} \, .  \label{e26}
\end{equation}
This gives the form of $\chi$ in the entire complex plane, where
$P_n$ is yet to be determined. Note that for this potential the
combination $a+b$ and $a-b$ are both odd {\it i.e}, 3 and 1
respectively. Hence, we use table 3 to obtain all the information
regarding the residues at the fixed poles, number of moving poles
of $\chi$, number of linearly independent solutions and their
form, for each set, by taking the values of $a=2$, $b=1$,
$M^{\prime}=0$ and $N^{\prime}=1$. The unknown polynomial in the
wave function can be obtained from \eqref{e23}, where $G(t)$ for
this potential satisfies,
\begin{eqnarray}
G(t) =\frac{t^2 (4b_{1}^{2}-2b_1 +\frac{1}{4})-2b_1
+\frac{1}{2}}{(t^2 -1)^2} + \frac{m^2 t^2 (4d_1 ^2 -2d_1
+\frac{1}{4})- 2md_1 - \frac{3m}{2}}{(mt^2 -1)^2} +  \nonumber \\
\frac{2E +8m +(16b_1 d_1 -13)mt^2}{2(1-t^2)(1-mt^2)}
\label{e27}
\end{eqnarray}
Using \eqref{e23} and \eqref{e27}, one gets the explicit
expressions for the eigenfunctions and the eigenvalues as given in
table 4. From the table, we see that the first set of residues
gives $n = -1$, which will not be considered as $n$ cannot be
negative. Thus  this particular case of Lam{\'e} potential has
five band edge solutions, which can be obtained analytically out
of an infinite number of possible states.

\section{ ALP with \protect\($a,\,b$ \protect\) half integers :}
The potential studied here is the supersymmetric associated
Lam{\'e} potential, with $a = 7/2$,  $b= 1/2$ :
\begin{equation}
V_{-} = \frac{63}{4}m\, \sn ^2 x +\frac{3}{4}m\frac{\cn ^2 x}{\dn
^2 x} -2 - \frac{29}{4}m + \delta_9     \label{e28}
\end{equation}
where $\delta_9 = \sqrt{4-4m +25m^2}$. In terms of $\chi(t)$, the
QHJ equation takes the form,
\begin{equation}
\chi ^2 + \chi ^{\prime} +\frac{2+t^2}{4(t^2-1)^2}+\frac{m^2
t^2-m}{4(mt^2-1)^2}+\frac{4E+8+29m-4\delta_9-65mt^2}{4(t^2
-1)(mt^2-1)} =0  \, . \label{e29}
\end{equation}
Note that, for this case, $a+b = 4$  and $a-b=3$, these are even
and odd respectively. Hence, for such cases, one needs to use sets
1 and 3 from table 2 and sets 2 and 4 from table 3 in order to get
the four groups of the eigenfunctions. The solutions for this
potential are given in table 5.  We see that there is a degeneracy
in the band edge energy eigenvalue, $14-7m+\delta_9$. All the
solutions agree with the known results [\ref{khare}].

\section{conclusions}
   In conclusion, in this study, we have demonstrated the applicability of QHJ formalism to QES periodic
potentials. We have been successful in obtaining the quasi-exact
solvability conditions and band edge solutions for cases when both
$a$ and $b$ are integers or half-integers. The origin of the large
number of solvability conditions for the ALP case, comes out
naturally in QHJ approach. Interestingly, the singularity
structure of the QMF for the QES periodic potentials is similar to
that of the ES periodic potentials. This structure is completely
different from those of the polynomial potentials. The case where
one of the parameter $a$ and $b$ is an integer, the other being a
half-integer, requires further careful study and will be reported
elsewhere.

{\bf References}

\begin{enumerate}

\item{\label{Singh}} Singh V, Biswas S N and Datta K 1978 {\it{Phys.
Rev.}} D {\bf{18}} 1901;\\Razavy M 1980 {\it{Am. J. Phys.}}
{\bf{48}} 285, 1981 {\it{Phys. Lett.}} A {\bf{82}} 7;\\Znojil M
1983 {\it{Phys. Lett.}} A {\bf{13}} 1445.

\item{\label{Turb}} Turbiner A V and Ushveridze A G 1987 {\it{Phys.
Lett.}} A {\bf{126}} 181;\\ Kamran N and Olver P J 1990 {\it{J. Math. Anal. Appl.}} {\bf{145}} 342;\\
Gonz$\acute{a}$lez-L$\acute{o}$pez A, Kamran N and Olver P J 1991
{\it{J. Phys. A: Math. Gen}} {\bf{24}} 3995.

\item{\label{Shifm}} Shifman M A 1989 {\it{Int. J. Mod. Phys.}} A
{\bf{4}} 2897 and references therein.

\item{\label{Ushve}} Ushveridze A G 1994 {\it{Quasi-Exactly Solvable Models
in Quantum Mechanics}} (Bristol, Inst. of Physics Publishing) and
references therein.

\item {\label{geo}} Geogo K G, Sree Ranjani S and Kapoor A K 2003 {\it{J. Phys
  A : Math. Gen.}} {\bf 36} 4591.

\item{\label{Atre}} Atre R and Panigrahi P K 2003 {\it{Phys.
Lett.}} A {\bf{317}} 46.

\item {\label{han}}Hancock H 1958 {\it{Theory of Elliptic Functions}} (New York, Dover
  Publications).

\item {\label{khare}} Khare A and Sukhatme U 1999 {\it{J. Math. Phys.}} {\bf
  40} 5473.

\item {\label{suk}} Khare A and Sukhatme U 2001 {\it{J. Math. Phys: Math. Gen.}} {\bf 42}
5652.

\item {\label{mag}} Magnus W and Wrinkler S 1966 {\it Hills
Equation} (New York, Inter Science Publishers).


\item {\label{asis}} Ganguly A 2002 {\it{J. Math. Phys.}} {\bf 43} 1980.

\item {\label{tka}} Tkachuk V M and Voznyak O 2002{\it{Phys. Lett.}} A {\bf301} 177.

\item {\label{lea}} Leacock R A and Padgett M J 1983 {\it{Phys. Rev. Lett.}} {\bf 50}
873.

\item{\label{pad}} Leacock R A and Padgett M J 1983 {\it{Phys. Rev.}} D {\bf 28}
2491.

\item {\label{sree}} Sree Ranjani S, Geojo K G, Kapoor A K and
Panigrahi P K, 2003 quant-ph/0211168.

\item {\label{akk}} Sree Ranjani S, Kapoor A K and Panigrahi P K, 2003
  quant-ph/0312041.

\end{enumerate}

\input{ntab1.tex}
\input{ntab2.tex}
\input{ntab3.tex}
\input{ntab4.tex}
\input{ntab5.tex}
\end{document}

%% file: ntab1.tex
\begin{table}[p]
\caption{The quasi exact solvability condition from the four
  permitted combinations of $b_1$ and $d_1$ for the general associated
  Lam{\'e } potential.}
\begin{tabular}{|c|c|c|c|c|}
\hline
\multicolumn{1}{|c|}{set} &\multicolumn{1}{|c|}{$b_1$}
&\multicolumn{1}{|c|}{$d_1$} &\multicolumn{1}{|c|}{$2b_{1} +2d_{1}
+n = \lambda_{1}$} &\multicolumn{1}{|c|}{QES condition}
\\ \hline
$1$ & $3/4$ & $\frac{3}{4}+\frac{b}{2}$ & $2+b+n = a$ & b-a = -n -2  \\
$2$ & $3/4$ & $\frac{1}{4}-\frac{b}{2}$ & $1-b+n = a$ & a+b+1 = n+2  \\
$3$ & $1/4$ & $\frac{3}{4}+\frac{b}{2}$ & $1+b+n = a$ & b-a = -n-1   \\
$4$ & $1/4$ & $\frac{1}{4}-\frac{b}{2}$ & $-b+n = a$  & a+b = n  \\
\hline
\end{tabular}
\end{table}

%% file: ntab2.tex
\begin{table}[p]
\caption{The form of the wave functions for the four sets of
residue combinations when $a+b$ and $a-b$ are even and equal to
$2N$ and $2M$ respectively.}
\begin{tabular}{|c|c|c|c|c|c|c|}
\hline
\multicolumn{1}{|c|}{set}
 &\multicolumn{1}{|c|}{$b_1$}
&\multicolumn{1}{|c|}{$d_1$}
 &\multicolumn{1}{|c|}{$n =\lambda_{1}- 2b_{1} -2d_{1}$}
&\multicolumn{1}{|c|}{$n(M,N)$} 
&\multicolumn{1}{|c|}{wave function $\psi(x)$}
&\multicolumn{1}{|c|}{ LI solutions}
\\ \hline
$1$ & $3/4$ & $\frac{3}{4}+\frac{b}{2}$ & $a-b-2$ & $2M-2$ & $\cn x (\dn x)^{1+b}P_{2M-2}(\sn x)$    & $M$ \\
$2$ & $3/4$ & $\frac{1}{4}-\frac{b}{2}$ & $a+b-1$ & $2N-1$ & $\frac{\cn x}{(\dn x)^b}P_{2N-1}(\sn x)$& $N$ \\
$3$ & $1/4$ & $\frac{3}{4}+\frac{b}{2}$ & $a-b-1$ & $2M-1$ & $(\dn x)^{b+1}P_{2M-1}(\sn x)$          & $M$ \\
$4$ & $1/4$ & $\frac{1}{4}-\frac{b}{2}$ & $a+b$   & $2N$   & $ \frac{P_{2N}(\sn x)}{(\dn x)^b}$      & $N+1$\\
\hline
\end{tabular}
\end{table}

%% file: ntab3.tex
\begin{table}[p]
\caption{The form of the wave functions for the four sets of
residue combinations when $a+b$ and $a-b$ are odd and equal to
$2N^{\prime} +1$ and $2M^{\prime}+1$ respectively.}
\begin{tabular}{|c|c|c|c|c|c|c|}
\hline
\multicolumn{1}{|c|}{set}
 &\multicolumn{1}{|c|}{$b_1$}
&\multicolumn{1}{|c|}{$d_1$}
 &\multicolumn{1}{|c|}{$n= \lambda_1 - 2b_1 - 2d_1$}
&\multicolumn{1}{|c|}{$n(M^{\prime},N^{\prime})$}
&\multicolumn{1}{|c|}{wave function $\psi(x)$ }
&\multicolumn{1}{|c|}{ LI solutions }
\\ \hline
$1$ & $3/4$ & $\frac{3}{4}+\frac{b}{2}$ & $a-b-2$ & $2M^{\prime}-1$ & $\cn x (\dn x)^{1+b}P_{2M^{\prime}-1}(\sn x)$    & $M^{\prime}$ \\
$2$ & $3/4$ & $\frac{1}{4}-\frac{b}{2}$ & $a+b-1$ & $2N^{\prime}  $ & $\frac{\cn x}{(\dn x)^b}P_{2N^{\prime}}(\sn x)$  & $N^{\prime}+1$ \\
$3$ & $1/4$ & $\frac{3}{4}+\frac{b}{2}$ & $a-b-1$ & $2M^{\prime}  $ & $(\dn x)^{b+1}P_{2M^{\prime}}(\sn x)$            & $M^{\prime}+1$ \\
$4$ & $1/4$ & $\frac{1}{4}-\frac{b}{2}$ & $a+b$   & $2N^{\prime}+1$ & $ \frac{P_{2N^{\prime}+1}(\sn x)}{(\dn x)^b}$    & $N^{\prime}+1$\\
\hline
\end{tabular}
\end{table}

%% file: ntab4.tex
\begin{table}[p]
\caption{For the  Associated Lam{\'e} potential $ V_- (x) = 6m
\,{\sn(x)}^2 + 2m \frac{{\cn(x)}^2}{{\dn(x)}^2} -4m$, with $a=2$
and $b=1$ the residues, the value of $n$, number of linear
independent solutions, the band edge eigenfunctions and
eigenvalues are as follows. Here $a+b =3$ and $a-b =1$ which give
$N^{\prime} =1$ and $M^{\prime}= 0$.}
\begin{tabular}{|c|c|c| c c c|c|c|c|}
\hline
\multicolumn{1}{|c|}{set}
 &\multicolumn{1}{|c|}{$b_1$}
&\multicolumn{1}{|c|}{$d_1$}
&\multicolumn{1}{|c}{    }
 &\multicolumn{1}{c}{$n$}
&\multicolumn{1}{c}{    }
&\multicolumn{1}{|c|}{LI solutions }
&\multicolumn{1}{|c|}{eigenfunction
$\psi(x)$ } &\multicolumn{1}{|c|}{eigenvalues}
\\ \hline
$1$ & $3/4$ & $5/4$  & & -1 & & - &                                                -      &        -                 \\
$2$ & $3/4$ & $-1/4$ & & 2 &  & 2 & $\frac{\cn x}{\dn x}(3m\sn ^2 x -2 \pm \sqrt{4-3m})$  &$ 5-3m \pm 2\sqrt{4-3m}$  \\
$3$ & $1/4$ & $5/4$  &  & 0  & & 1 & $\dn ^2 x                                $            & 0                        \\
$4$ & $1/4$ & $-1/4$ & & 2 &  & 2 & $ \frac{\sn x}{\dn x}(3m\sn ^2 x-2-m\pm \sqrt{4-5m+m^2}$ & $5-2m \pm 2\sqrt{m^2 -5m +4}$\\
\hline
\end{tabular}
\end{table}

%% file: ntab5.tex
\begin{table}[p]
\caption{For the  Associated Lam{\'e} potential $ V_- (x) =
\frac{63}{4}m\, \sn ^2 x +\frac{3}{4}m\frac{\cn ^2 x}{\dn ^2 x} -2
- \frac{29}{4}m + \delta_9 $ with $a=7/2$ and $b=1/2$ the
residues, the value of $n$, number of linear independent
solutions, the band edge eigenfunctions and eigenvalues are as
follows. Here $a+b =4$ and $a-b =3$ which give $N =2$ and
$M^{\prime}= 1$.}
\begin{tabular}{|c|c|c| c c c|c|c|c|}
\hline
\multicolumn{1}{|c|}{set}
 &\multicolumn{1}{|c|}{$b_1$}
&\multicolumn{1}{|c|}{$d_1$}
&\multicolumn{1}{|c}{     }
 &\multicolumn{1}{c}{$n$}
&\multicolumn{1}{c}{     }
&\multicolumn{1}{|c|}{ LI solutions }
&\multicolumn{1}{|c|}{eigenfunction $\psi(x)$ }
&\multicolumn{1}{|c|}{eigenvalues}
\\ \hline
 1  & $3/4$ & 1 & & 1 & & 1 &  $\cn x (\dn x)^{3/2}\sn x$                & $\delta_9 -m +2$  \\
    &       &   & &   & &   &                                 &   \\
 2  & $3/4$ & 0 & & 3 & & 2 &  $\cn x (\dn x)^{3/2}\sn x $               & $\delta_9 -m +2 $ \\
    &       &   & &   & &   &  $\cn x (\dn x)^{-1/2}\sn x (1-2\sn ^2 x)$ &$14 -7m+\delta_9$  \\
    &       &   & &   & &   &               &   \\
 3  & $1/4$ & 1 & & 2 & & 2 &$(\dn x)^{3/2}(12m\,\sn ^2x-5m-2-\delta_9)$ & 0               \\
    &       &   &   & & &   &$(\dn x)^{3/2}(12m\,\sn ^2x-5m-2+\delta_9)$ &$ 2\delta_9$      \\
    &       &   &   & & & &  &  \\
 4  & $1/4$ & 0 & & 4 & & 3 &$(\dn x)^{3/2}(12m\,\sn ^2x-5m-2-\delta_9)$ & 0               \\
    &       &   & &   & &   &$(\dn x)^{3/2}(12m\,\sn ^2x-5m-2+\delta_9)$ & $2\delta_9$  \\
    &       &   & &   & &   & $1-8\sn ^2 x\cn^2 x $                      & $14-7m +\delta_9$    \\
\hline
\end{tabular}
\end{table}